%% file: main.tex
\documentclass[copyright,creativecommons]{eptcs}

\usepackage{iftex}

\usepackage{tikzit}
\input{basic.tikzdefs}
\input{basic.tikzstyles}

\usepackage[utf8]{inputenc}
\usepackage{graphicx}

\usepackage{amsmath}
\usepackage{amsthm}
\usepackage{amssymb}

\newtheorem{prop}{Proposition}

\theoremstyle{definition}
\newtheorem{defn}{Definition}
\newtheorem{example}{Example}

\title{Identification of Causal Influences in \\ Quantum Processes}
\author{Isaac Friend \qquad\qquad Aleks Kissinger
\institute{Department of Computer Science\\
University of Oxford}
\email{\{isaac.friend,aleks.kissinger\}@cs.ox.ac.uk}
}
\def\titlerunning{Quantum Causal Identification}
\def\authorrunning{I. Friend \& A. Kissinger}

\begin{document}

\maketitle

\begin{abstract}
Though the topic of causal inference is typically considered in the context of classical statistical models, recent years have seen great interest in extending causal inference techniques to quantum and generalized theories.  Causal identification is a type of causal inference problem concerned with recovering from observational data and qualitative assumptions the causal mechanisms generating the data, and hence the effects of hypothetical interventions.  A major obstacle to a theory of causal identification in the quantum setting is the question of what should play the role of ``observational data,'' as any means of extracting data at a locus will almost certainly disturb the system. Hence, one might think \textit{a priori} that quantum measurements are already too much like interventions, so that the problem of causal identification trivializes. This is not the case.  Fixing a limited class of quantum instruments (namely the class of all projective measurements) to play the role of ``observations,'' we note that as in the classical setting, there exist scenarios for which causal identification is impossible.  We then present sufficient conditions for quantum causal identification, starting with a quantum analogue of the well-known ``front-door criterion'' and finishing with a broader class of scenarios for which the effect of a single intervention is identifiable.  These results emerge from generalizing the process-theoretic account of classical causal inference due to Jacobs, Kissinger, and Zanasi beyond the setting of Markov categories, and thereby treating the classical and quantum problems uniformly.
\end{abstract}

\section{Introduction}
The problem of causal inference is to deduce from statistical correlations among variables something about the causal mechanisms responsible for those correlations, where a causal mechanism is a process that answers \textit{interventional} queries. Although the majority of the work in the field of causal inference has focused on classical, statistical models, it is interesting to consider causal inference problems in the quantum setting as well, where quantum systems play the role of classical random variables. One can ask, for example, whether it is possible for agents confronted with a recurring scenario involving a pair of quantum systems to deduce using only certain limited operations whether the agents are in a common-cause-type situation (e.g., accessing two parts of a quantum entangled state) or a cause-effect-type situation (e.g., accessing the same system at two points in time).  Ried et al.  \cite{ried_quantum_2015} presented a solution to such an inference problem for specific scenarios involving two quantum systems, and raised the problem of inference with larger collections of systems.  Essentially what is sought is quantum generalization of some of the theory of statistical causal inference, which has systematized much of the business of combining qualitative knowledge of ``causal structure" with quantitative data to characterize causal influences between variables.  This article accomplishes such generalization, using the logical conception of causality presented in \cite{jacobs_causal_2019} to reveal the common process-theoretic underpinnings of causal inference in both ordinary stochastic and quantum settings.

A theory of quantum causal inference requires first a mathematical model of quantum causal scenarios. Here, we will take a minimal notion of a quantum causal model consisting of a ``circuit with holes,'' i.e., a directed acyclic string diagram wherein some wires have gaps allowing agents to apply local processes. The circuit can be seen as a second-order process, or \textit{comb} \cite{chiribella_quantum_2008}, which maps local non-deterministic processes to probabilities. 

This notion of causal model is relatively weak in that unlike the one studied in \cite{allen_quantum_2017} and ~\cite{barrett_quantum_2019}, it doesn't seem to admit a relation of ``complete common cause'' whereby a single intervenable quantum system can act as the sole source of correlations between two systems in its future.  As a complete common cause can be pictured in the classical setting as ``copying'' a random variable and using it as input to two or more subsequent stochastic maps, it is difficult and somewhat subtle to make sense of a ``complete quantum common cause'' in the absence of a physically meaningful process of cloning or broadcasting quantum systems. Hence, it is interesting to see how much traction we can get on causal inference for a class of models that don't admit the explicit general representation of complete common causes. We will show here that, in the case of the particular problem of \textit{quantum causal identification}, we can get relatively far without such a representation.  We also recover, from an abstract perspective, results in classical statistical causal inference.

Causal identification, in the classical case, refers to the problem of identifying the effects of (often hypothetical) interventions on the basis of purely observational data~\cite{pearl_causality_2009}. In contrast to related problems such as causal discovery, here the hypothesized causal structure of events--represented, e.g., by a directed acyclic graph depicting the possible directions of causal influence between random variables--is known in advance, but not the exact conditional probability distributions (or functional dependencies) governing the influence of individual variables on each other.  Even with the causal structure given in advance, this problem can be highly non-trivial in the presence of confounding variables~\cite{pearl_causality_2009} or selection bias~\cite{correa_identification_2019}.

In generalizing to quantum causal identification, one needs to fix a notion that stands in the place of ``observation,'' as it is impossible to extract any data from a quantum system without causing a disturbance, which in some sense is already an active intervention. Here, we fix the class of processes playing the role of ``observations'' as local projective measurements, whereas ``interventions'' can be arbitrary quantum instruments. The latter includes, for example, the process of discarding the incoming state of a system and preparing a fixed new state, while the former does not.

While we do not intend to argue here that these notions of ``observation'' and ``intervention'' are fully conceptually justified, we will give strong evidence instead that this kind of quantum causal identification problem is interesting: we note that the problem can be impossible, then show that it becomes possible when a causal structure satisfies certain criteria.

By analogy to the classical case, causal identification is defined to be impossible when a causal structure admits a pair of models that behave identically with respect to projective measurements, but differently under arbitrary interventions.  Simple such pairs of models were mentioned in \cite{ried_quantum_2015}, and we give an example.  Our first new result is a quantum version of the front-door criterion for causal identifiability \cite{pearl_causal_1995}.  This result is then generalized to a sufficient condition for identifiability that implies the quantum analogues of multiple sufficient conditions in the statistical causal modeling literature, including some cases covered by Galles and Pearl in \cite{galles_testing_1995} and by Tian and Pearl in \cite{tian_general_2002}.  The statements and proofs here invoke diagrammatic technology presented in \cite{coecke_picturing_2017} and previously applied to causal inference by Jacobs, Kissinger, and Zanasi \cite{jacobs_causal_2019}, who indicated the possibility, realized in the present article, of ``import[ing] results from classical causal reasoning to the quantum case" by changing the concrete process theory in which abstract causal diagrams are modeled.

Potential consequences of the work lie in multiple areas.  With respect to applicable quantum information science, the present work first of all describes how to identify certain interventional quantities in quantum networks of certain shapes, without full tomography.  In fact, our results indicate that the abstract causal structure of a collection of quantum processes can sometimes be used to characterize those processes completely, even with limited interactions.  As Ried et al. explained, quantum causal inference schemes with limited operations ``promise extensive applications in experiments exhibiting quantum effects'' \cite{ried_quantum_2015}.  Our means of inference in scenarios involving unobserved common causes might apply specifically to the problem of detecting non-Markovianity in quantum information processing \cite{rivas_quantum_2014}.

This work may also have consequences for the theory of non-quantum statistical causal inference.  The process-theoretic presentation here, unifying classical and quantum causal identification, uncovers the basic structures and procedures--comb factorization, informationally complete sets of states and effects, and process tomography--that underpin causal inference but are often masked by the details of classical probability theory.  The isolation of these rudiments should not only help guide the further development of theories of causal inference for quantum and other special kinds of processes, but also motivate continued research in ordinary statistical causal modeling using the logical and compositional techniques of theoretical computer science.

Finally, we hope this work will contribute to the program aimed at answering questions in the foundations of quantum physics by viewing them through the lens of causal modeling and inference \cite{wood_lesson_2015}.  In order to draw foundational conclusions from our results, those pursuing such ideas will have to assess the implications of the fact that projective measurement as ``passive observation" defines a close quantum analogue of the classical problem of causal identification with latent variable models.

\section{Preliminaries}

To treat classical probability and quantum theory on the same footing, we will use the language of \textit{process theories} \cite{coecke_picturing_2017} throughout.  Process theories have been defined in slightly varying ways in the literature. Our definition follows.
\begin{defn}\label{def:process-theory}
A \textit{process theory} is a symmetric monoidal category $(\mathcal C, \otimes, I)$.
\end{defn}
The concrete classical and quantum process theories of causal models studied in this work are each equipped with a distinguished family of \textit{discarding} morphisms $d_A : A \to I$ for each object $A$, satisfying $d_{A\otimes B} = d_A \otimes d_B$ and $d_I = 1_I$.

To give a physical or computational interpretation to process theories, it is typical to refer to generic morphisms $f : A \to B$ as \textit{processes}, morphisms of the form $\rho : I \to A$ as \textit{states}, and morphisms of the form $\pi : A \to I$ as \textit{effects}. Morphisms of the form $\lambda : I \to I$ are called \textit{numbers} or \textit{scalars}.  Objects are also called \textit{system-types}.  

Throughout the paper, we will adopt \textit{string diagram} notation, where processes are depicted as boxes and objects as wires. We depict discarding using a black dot.
\ctikzfig{string-diagrams}

Note that the discarding maps in a process theory are not required \textit{a priori} to  satisfy any equations aside from the basic compatibility with $\otimes$. They play an important role, however, in identifying certain families of well-behaved maps within a process theory. The most important such condition is the following. 

\begin{defn}\label{def:causal}
A map $f : A \to B$ is called \textit{causal} if $d_B \circ f = d_A$, or diagrammatically:
\begin{equation}\label{eq:causal}
    \input{causal.tikz}
\end{equation}
\end{defn}
Intuitively, causality captures the fact that the only influence a map can have is on its ``future,'' i.e., its output. If the output is discarded, then the actual causal process that took place is irrelevant.

Our main examples of process theories are $\textbf{Mat}[\mathbb R_+]$ and $\textbf{CPM}$, which contain (finite-dimensional) classical probability theory and quantum theory, respectively.

\begin{example}\label{ex:stoch}
The process theory $\textbf{Mat}[\mathbb R_+]$ has as objects natural numbers and as morphisms $M : m \to n$ the $n \times m$ matrices whose entries are non-negative real numbers. The monoidal product is given by tensor product of matrices (a.k.a. Kronecker product), whose unit is the $1\times 1$ matrix $(1) : 1 \to 1$. Discarding maps $d_n : n \to 1$ are the $1 \times n$ matrices (i.e. row vectors) consisting of all $1$'s. Consequently, causal states are column vectors of positive numbers whose entries sum to 1 (i.e., probability distributions), and causal processes are matrices whose columns each sum to 1 (i.e., stochastic maps, equivalent to conditional probability distributions with $P(i|j) := M_{ij}$).
\end{example}

\begin{example}\label{ex:cpm}
The process theory $\textbf{CPM}$ has as objects finite-dimensional Hilbert spaces $\mathcal H, \mathcal K, ...$ and as morphisms completely positive maps $\Phi : L(\mathcal H) \to L(\mathcal K)$, where $L(\mathcal H)$ is the algebra of operators $\mathcal H \to \mathcal H$. The monoidal product is again given by tensor product, whose unit is the identity map on $L(\mathbb C) \cong \mathbb C$. States $\rho : \mathbb C \to L(\mathcal H)$ are fixed by a single positive operator $\rho(1) \in L(\mathcal H)$ and causal states correspond to trace-1 positive operators. More generally, causal processes are the trace-preserving completely positive maps.
\end{example}

We will furthermore find it convenient to assume that each process theory has a (self-dual) compact structure, meaning that every object $A$ is equipped with a pair of maps $\cup_A : I \to A \otimes A$ and $\cap_A: A \otimes A \to I$, called ``cups'' and ``caps'' respectively, satisfying the so-called \textit{yanking equations}, which are depicted in string diagram notation as follows:
\[
\input{line_yank.tikz}
\]

This structure enables us easily to represent higher-order maps as first order ones. For example, we can represent a process that takes processes of type $A \to A'$ and produces processes of type $B \to B'$ as a normal, first-order process $f: B \otimes A' \to A \otimes B'$. We then indicate its higher-order interpretation by drawing $f$ as a box with a ``hole'' in it, and use cups and caps to define ``plugging'' another box into that hole:
\noindent\begin{minipage}{.4\linewidth}
\begin{equation}\label{eq:2ord-comb}
\input{2-comb-ch.tikz} \leadsto\;\;
\input{2-comb2.tikz}
\end{equation}
\end{minipage}%
\begin{minipage}{.6\linewidth}
\begin{equation}\label{eq:2-comb-compose}
\input{2-comb-compose.tikz}
\end{equation}
\end{minipage}

In \cite{jacobs_causal_2019}, the authors furthermore assumed the structure of a CDU category--a minor variation on the notion of a Markov category~\cite{fritz_synthetic_2020}--which captures an abstract notion of probabilistic maps by assuming every object carries a ``copying'' (a.k.a. ``broadcasting'') map~\cite{coecke_picturing_2012}. In particular, this structure allows one to capture causal models based on Bayesian networks as certain functors between CDU categories.

The famous no-cloning/no-broadcasting theorems of quantum theory, however, rule out a Markov-like structure in the category \textbf{CPM} of quantum maps. Hence, we adopt a weaker notion of causal model, consisting of a formal string diagram (i.e., a morphism in the free category over a signature) and an interpretation of that diagram into a concrete process theory (of, e.g., probabilistic or quantum maps).

\section{Interventional causal models}

A causal model consists of two parts: (i) a formal string diagram capturing our causal hypotheses, and (ii) an associated interpretation in a concrete process theory (i.e., $\textbf{Mat}[\mathbb R_+]$ or \textbf{CPM}).  We will also use the word ``model" to refer just to (ii): the interpretation in the concrete process theory gives a model of (i) in a logical sense.

We define a formal string diagram as a morphism of a particular form in the free symmetric monoidal category $\textbf{Free}(\Sigma)$ over some signature $\Sigma$. For a fixed set of objects $\{X_1, \ldots, X_n \}$ in $\Sigma$, we call a diagram $D: X_1 \otimes \ldots \otimes X_n \to X_1 \otimes \ldots \otimes X_n$ a \textit{circuit with holes} if it is a morphism in the free symmetric monoidal category and furthermore has the property that joining each input $X_i$ to its corresponding output $X_i$ yields another morphism in the free SMC (i.e., it doesn't introduce a directed cycle).

The intuition is that each of the input/output pairs is a ``hole'' in the diagram, which we call an \textit{intervention locus}, or simply \textit{locus} (plural loci), where a local process can be plugged in. For example:
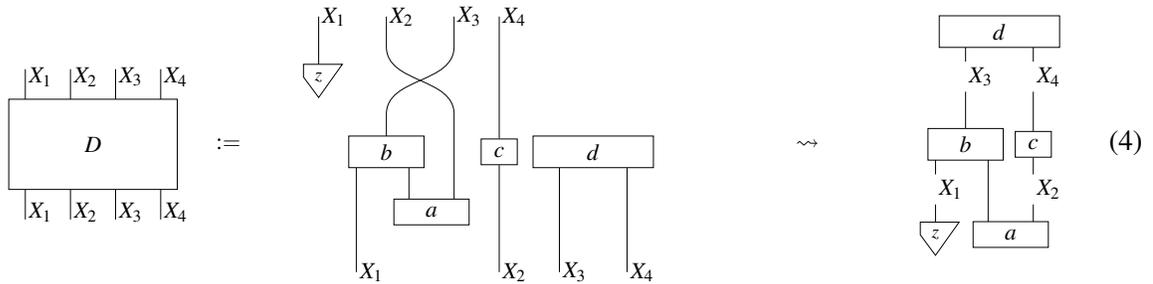
\begin{equation}\label{eq:causal-model}
    \input{causal-model.tikz}
\end{equation}
We require a locus's input and output system-types to be identical, partly in order to accommodate the special ``trivial intervention," which joins a locus's input and output with an identity wire.  More broadly, statistical causal inference often involves considering a pair of instances of a single variable, with, e.g., one being observed and the other set by intervention.  What is being represented is the same causal relatum at two different ``times," before and after intervention.  

We can now introduce a notion of causal model that is similar in spirit to that of~\cite{jacobs_causal_2019}, but no longer relies on the CDU structure needed to capture Bayesian networks.

\begin{defn}\label{def:causal-model}
For any process theory $\mathcal C$, an interventional causal model consists of a pair $(D, \Phi)$ where $D$ is a circuit-with-holes in $\textbf{Free}(\Sigma)$, $\Phi$ is a causal process in $\mathcal C$, and there exists a symmetric monoidal functor $F: \textbf{Free}(\Sigma) \to \mathcal C$ such that $F(D) = \Phi$.
\end{defn}
When $\mathcal C = \textbf{Mat}[\mathbb R^+]$ we call $(D, \Phi)$ a \textit{classical interventional causal model}, whereas when $\mathcal C = \textbf{CPM}$, we call it a \textit{quantum interventional causal model}.

The shape of the abstract diagram $D$ containing loci $X_i$ and $X_j$ may prohibit models in which interventions at $X_i$ can causally affect events at $X_j$.  Call locus $X_j$ a \textit{descendant} of locus $X_i$ in abstract diagram $D$ if and only if, when every input other than $X_i$ and $X_j$ is joined to its corresponding output, the resulting circuit has a path from input $X_i$ (the wire leaving locus $X_i$) to output $X_j$ (the wire arriving at locus $X_j$) along which every traversal of a box is in the upward direction.  Thus in the example depicted in \ref{eq:causal-model}, $X_4$ is a descendant of $X_2$, but not of $X_3$.  If locus $X_j$ is not a descendant of locus $X_i$ in abstract diagram $D$, then for any model of $D$, for every fixed set of causal processes plugged into the loci other than $X_i$ and $X_j$, plugging a causal process into locus $X_i$ yields a process, with only input/output pair $X_j$, that does not depend on the choice of causal process at $X_i$.  In other words, $X_j$ being a non-descendant of $X_i$ in $D$ captures the hypothesis that interventions at locus $X_i$ cannot possibly affect events at $X_j$, and the problem of identifying the causal influence of $X_i$ on $X_j$ is uninteresting.  Note finally that if $X_i$ is a descendant of $X_j$, then $X_j$ is not a descendant of $X_i$, and the uninteresting identification scenario obtains.  We therefore lose nothing by focusing henceforth on the case wherein $X_i$ is not a descendant of $X_j$.

Inferring the ``causal influence" of one locus on another will mean inferring, from whatever data are available, the value of a certain process determined by the causal model.  That process, called an \textit{interventional channel} between the two loci, encodes the quantitative causal relation between the loci according to the model at hand.  
\begin{defn}
In a classical or quantum interventional causal model with loci $X_1$, \ldots, $X_n$, the \textit{interventional channel from $X_i$ to $X_j$}, where $X_i$ is a non-descendant of $X_j$, is the process obtained by filling in all loci other than $X_i$ and $X_j$ with identity interventions, and inputting a normalized, i.e., causal, state to the wire leaving locus $X_j$.
\end{defn}
The interventional channel--whose definition as above using an unspecified causal state is made possible by the assumption that $X_i$ is a non-descendant of $X_j$--is a process of the form
\ctikzfig{causal-effect}
which maps (possibly non-deterministic) intervention outcome--$f: X_i \rightarrow X_i$ at locus $X_i$ to the state on system $X_j$ resulting from the combination of intervention $f$ at $X_i$ and trivial (identity) interventions at all loci other than $X_i$ and $X_j$.  In particular, the interventional channel gives the consequence for $X_j$ of forcibly setting the state leaving $X_i$ to $\psi$:
\ctikzfig{do-effect}
Thus the interventional channel yields what in ordinary statistical causal inference is called an ``interventional distribution" of $X_j$ due to ``surgical intervention" at $X_i$.  Moreover, one can compose the interventional channel with arbitrary causal processes $X_i \rightarrow X_i$ to evaluate the influences of so-called soft interventions \cite{correa_calculus_2020}, for which the state leaving a locus depends on the incoming state.    

Thus the shift in focus from distributions to channels, in line with a general trend toward channel-based accounts of probabilistic reasoning \cite{jacobs_logical_2018, cho_disintegration_2019} and suggested for the present work by the difficulties of defining and reasoning with quantum analogues of conditional probability distributions, has definite advantages for a theory of causal inference.  The single process called the interventional channel supports uniform reasoning about the consequences of all kinds of interventions, including soft interventions, which are likely to be the norm in applications to quantum information processing, where the interventions under consideration may be, e.g., coherent quantum processes.  The channel-based approach is both conceptually clarifying in its application to classical causal models, and especially suited for the most pertinent problems of quantum causal inference.  

One expects that an interventional causal model's data-generating process should be completely discoverable from the results of various interventions \cite{costa_quantum_2016}.  That is, there should exist a set of instruments for which the outcome statistics suffice to determine the data-generating process.  This property is guaranteed for the interventional causal models in the present article by a key commonality between the classical and quantum processes studied in this work: they can be completely specified by the numbers that result when they are locally composed with states and effects.
\begin{prop}
The theories $\textbf{Mat}[\mathbb R_+]$ and $\textbf{CPM}$ have local process tomography: any process 
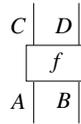
\ctikzfig{fABCD}
is determined by numbers
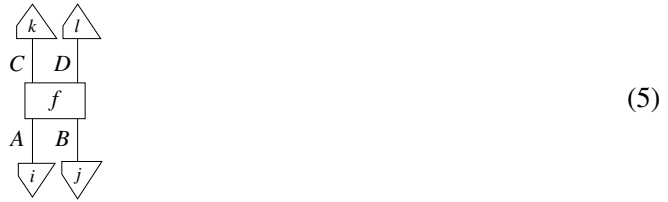
\begin{equation}\label{eq:matrix-elem}
\input{loc-proc-tomog.tikz}
\end{equation}
where $i, j, k,$ and $l$ index certain finite sets of states or effects on the appropriate system-types.  In quantum tomography literature, the appropriate sets are called ``informationally complete." 
\end{prop}

We call the set of numbers in equation~\eqref{eq:matrix-elem} the \textit{generalized matrix elements} associated with a process $f$.  A local process tomography protocol for causal--i.e., probability-preserving--maps in $\textbf{Mat}[\mathbb R_+]$ and $\textbf{CPM}$ uses observed probabilities of combinations of measurement outcomes conditioned on combinations of causal state preparations.  In the quantum case, though one cannot obtain all of the generalized matrix elements using a single choice of measurement basis, it is always possible to obtain them from the measurement statistics of multiple projective measurements at the outputs (along with independent state preparations at the inputs).    

Local process tomography for comb-shaped quantum processes, which corresponds to Ried et al.'s \cite{ried_quantum_2015} ``causal tomography,'' is mathematically just the same as local process tomography for ordinary first-order processes, but in the physical implementation, the measurement realizing an effect at a locus precedes temporally the preparation of the state leaving that locus.  Thus local process tomography for a classical or quantum interventional model typically relies on probabilities that result from filling intervention loci with maps of the form
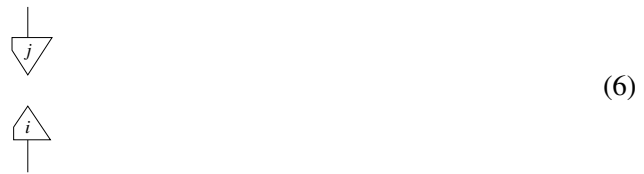
\begin{equation} \label{int-outcome}
\input{int-outcome.tikz}
\end{equation}
where $i$ and $j$ index informationally complete sets of effects and causal states.  To implement all these maps in an experiment and thereby learn corresponding probabilities requires the ability to record an observation labeled $i$ and then prepare the system in a new state labeled $j$, where $j$ does not depend on $i$.  

In contrast, what we will call observational data arise, for instance, when the only outcomes of this form that can be implemented are those satisfying $i = j$: the state to be fed forward from a locus is determined by the observation outcome, and so, for simplicity, we label them with the same value of a single index.  This article is concerned with what can be inferred circumstances like these.  The general problem of causal identification is to use qualitative assumptions about the causal scenario to compute quantitative causal influences given statistics from only a highly restricted set of interventions.  Usually the allowed interventions are ``passive observations," which non-deterministically implement certain maps of the aforementioned ``observational" form, and thereby teach the observer certain limited sets of probabilities.  There is no quantum instrument representing a procedure appropriately called passive observation. For the purposes of this paper, the quantum interventions allowed as ``observations" are exactly the projective measurements, which include identity processes (totally uninformative measurements) as well as instruments whose outcomes take the form of \eqref{int-outcome} for $i=j$.  

The class of projective measurement instruments is closely related to a criterion, called \textit{informational symmetry}, whereby Ried et al. characterized certain interventions in both classical and quantum causal scenarios as mere observations \cite{ried_quantum_2015}.  Informational symmetry, however, depends on both the intervening process and the prior state, whereas we desire a criterion applying only to the intervening process itself. 

To apply our proofs of sufficient conditions for identifiability to the classical stochastic setting, we need not characterize completely a classical stochastic analogue of the quantum class of observation outcomes, but only posit that classical observation outcomes include identity matrices and matrices that have all zero entries except $1$ in a single position on the main diagonal.  (When classical probability theory is viewed as a sub-theory of quantum theory, the non-identity classical observation outcomes just described are in fact identified with outcomes of maximally informative projective measurement in a fixed basis.)  The latter kind of matrix represents an outcome of what is normally called ``observing a random variable.''  By marginalization, identity interventions in the classical setting can be simulated from the probabilities of such projections onto pure causal states (point distributions).  Thus our proofs of classical identifiability really appeal to no intervention procedures other than ordinary maximally informative classical observation.  When we say an inference in the classical setting is impossible with only observational data accessible, ``observational data" means probabilities of the classical outcomes just described.  

The question of identifiability of the causal influence of one locus on another is whether the qualitative causal assumptions encoded in the abstract string diagram $D$ are strong enough that together with observational data for a causal scenario represented by an unknown model of $D$, they determine the value of the interventional channel derived from the unknown model. A classical or quantum interventional channel, respectively, from one locus to another will be called identifiable from an abstract string diagram if for any \textit{positive} stochastic or quantum model of the string diagram, the interventional channel can be computed from the probabilities of arbitrary combinations of observation outcomes at all intervention loci of the model.  Positivity is defined as follows:

\begin{defn}
A positive stochastic or quantum interventional model is a model whose composition with any non-zero state and any non-zero effect gives a strictly positive number.
\end{defn}

The states and effects composed with a model may in particular be products of those implemented at individual intervention loci.  For a positive model, therefore, any combination of observational outcomes occurs with non-zero probability.  The positivity condition in our process-theoretic account serves the same purpose as the common requirement in ordinary causal modeling that a probabilistic causal model induce a strictly positive joint distribution on all variables.  Positivity ensures that all relevant conditional probabilities are defined, and that detecting an arbitrary state at a locus after intervening at another locus is at least possible--if it were not, asking for the corresponding interventional probability would make no sense.  The definition of identifiability from an abstract string diagram captures the notion that the assumptions of no direct influence between loci disconnected in the abstract diagram--equivalent to assumptions of absence of certain arrows in a directed acyclic graph representing a classical \cite{pearl_causality_2009} or quantum \cite{costa_quantum_2016} causal structure--suffice for inference: in any model satisfying at least the constraints implied by the string diagram, the quantity in question can be deduced from observational outcome statistics.

Circumscribing the class of allowed interventions raises the question of whether the restrictions are strong enough to rule out schemes like causal tomography that would always allow causal identification.  The answer is affirmative, as Ried et al. \cite{ried_quantum_2015} noted, and is evident from string diagrams like
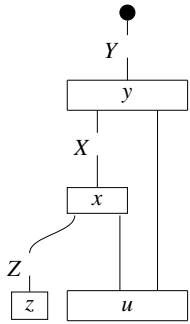
\ctikzfig{bad-model}
for which the interventional channel
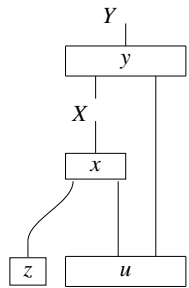
\ctikzfig{bad-model-effect}
from $X$ to $Y$ is not identifiable. Two models yielding different interventional channels but identical observational outcome statistics are constructed via functorial interpretation according to Definition \ref{def:causal-model}: in both models, $u$ is interpreted as the Bell state $|\Psi^+ \rangle \langle \Psi^+| = \frac{1}{2} (|0 \rangle + |1 \rangle) (\langle 0| + \langle 1|)$ on two qubits, $z$ as a fixed quantum state with full support (i.e., a state whose composition with any non-zero effect is non-zero, corresponding to an operator of full rank), and $x$ as the quantum map that discards its left-hand input and outputs its right-hand input unchanged.  In the first model, $y$ is interpreted as the map that discards its right-hand input and applies to its left-hand input a projective measurement followed by a depolarizing channel with parameter $\lambda$.  In the second model, $y$ is interpreted as the map that discards its left-hand input and applies to its right-hand input the same projective measurement followed by the same depolarizing channel as in the first model.  Thus the interpretations of $y$ in the two models are
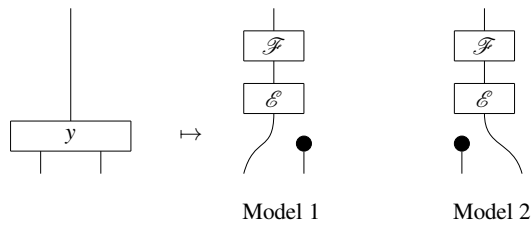
\ctikzfig{bad-model-y}
where
\begin{align*}
    \mathcal{E}(\rho) &= |0 \rangle \langle 0| \rho |0 \rangle \langle 0| + |1 \rangle \langle 1| \rho |1 \rangle \langle 1|\\
    \mathcal{F}(\sigma) &= (1 - \lambda)\sigma + \lambda (\frac{1}{2}|0 \rangle \langle 0| + \frac{1}{2}|1 \rangle \langle 1|).
\end{align*}

These models are positive because $z$ has full support, the reduced state arriving at $X$ is maximally mixed, and the state arriving at $Y$ includes a maximally mixed state with weight $\lambda$ no matter what outcome has occurred at $X$.  They produce identical outcome statistics for projective measurements at the loci.  The interventional channel is therefore not identifiable from the abstract string diagram, because it cannot be computed for \textit{every} positive model.  The theory behind the two-locus quantum identification schemes of \cite{ried_quantum_2015}, however, might sometimes help in three-locus situations--perhaps some scenarios that involve coherence or entanglement and also include an ``instrumental variable," which would correspond here to the locus $Z$.

Because $Z$ does not influence $X$ or $Y$ in these models, this example is essentially equivalent to those two-variable scenarios in \cite{ried_quantum_2015} for which the desired interventional channel was noted to be unidentifiable.  We show three variables to detach our example from the two-variable case of the ``quantum advantage."  Moreover, we explicitly note that identification is impossible even for positive models of the string diagram.

\section{Front-door scenarios}
It is generally impossible to tell from observational data whether two correlated random variables, one of which is known \textit{not} to be a descendant of the other--i.e., one of which comes ``before" the other--stand in a cause-effect relation or are instead descendants of an unobserved common cause.  If, however, there is a third observed variable or set of variables along the possible path of causal influence between the first two, the ``front-door criterion" for causal identifiability implies that such inference may be possible.  The operative sufficient condition has a quantum analogue, captured along with the classical version by the following result, which is derived for both process theories simultaneously, using the fact that the theories' scalars are real numbers. 

In this and the following section, each system represented by an uppercase letter may be a composite of multiple smaller systems, and similarly each box may be a composite of smaller boxes. Thus, a single locus in one of our diagrams might correspond to a list of several classical variables or quantum laboratories \cite{costa_quantum_2016} occupying several nodes of a more traditional causal diagram.  What we call an intervention at a locus would then correspond to (possibly choreographed/non-local) intervention at all those nodes.  

\begin{prop}\label{prop:front-door}
For quantum or stochastic models of a string diagram
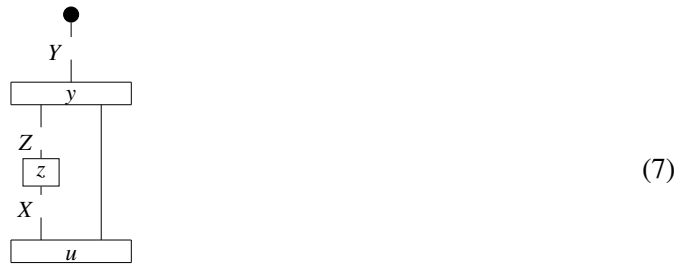
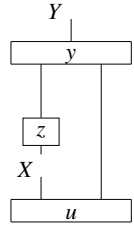
\begin{equation}\label{eq:front-door}
\input{front-door-model.tikz}
\end{equation}
the interventional channel
\ctikzfig{front-door-causal-effect}
from $X$ to $Y$ is identifiable.
\end{prop}
Each proof of identifiability from here on consists in demonstrating how to compute the causal quantity of interest--usually an interventional channel--from certain component processes of the model, specified by their generalized matrix elements, which are ultimately computed from probabilities of local observation outcomes.  Each proof can be read in either the classical or the quantum process theory.  An observation outcome consisting of an effect labeled, say, $i$, followed by a state with the same label, is to be understood in the classical case as a row vector with $1$ in the $i$th position followed by the transpose of that vector.  The composite map is the matrix product of the two.  In the quantum case, a state/effect pair represents a single outcome of a non-degenerate projective measurement. The state labeled $i$ is the $i$th measurement eigenstate, whereas the effect is the CPM obtained by tracing the input together with the $i$th eigenstate.
\begingroup
\allowdisplaybreaks
\begin{proof}
First, we compute the process $z$, determined by its generalized matrix elements, which we obtain by introducing a non-zero scalar factor and its inverse (where the inverse is indicated by a diagram inside ${\color{blue}\{{\color{black}-}\}^{-1}}$), then using the causality equation~\eqref{eq:causal} to transform into the following quantity:
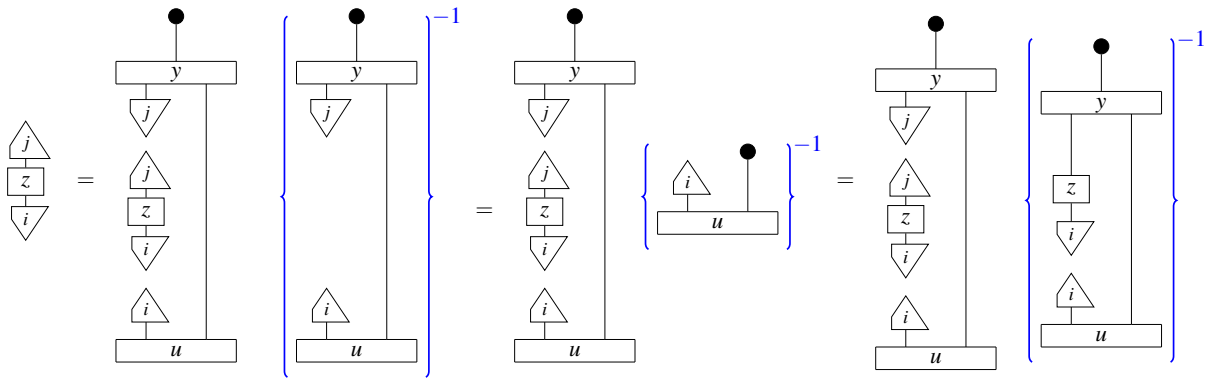
\[
    \input{z.tikz}
\]
Note that the scalars being inverted are indeed non-zero, by positivity of the whole interventional model. Furthermore, the rightmost diagram above consists of quantities that can be computed purely from projective measurements at all of the loci (including the identity/trivial measurement at $Z$).

Once we have computed the generalized matrix elements of $z$, we can use them to compute those of another factor of the model by a procedure we call ``adjusting for $z$:"
    \ctikzfig{fdstep2}
Finally, we compose the two processes at $Z$, leaving the $X$ input and output, to obtain the interventional channel.
\end{proof}
\endgroup


Thus, in the quantum just as in the classical case, observation at a locus $Z$ lying on the path between $X$ and $Y$ ``blocks" that path and allows control of the confounding influence of $u$.

\section{A more general case of a single intervention}
The identification criterion of Proposition \ref{prop:front-door} can be generalized, using the same proof technique, to a quantum version of Jacobs, Kissinger, and Zanasi's Theorem 8.1 \cite{jacobs_causal_2019}.

\begin{prop}\label{prop:thm8}
For quantum or stochastic models of a string diagram
\ctikzfig{thm-8-model}
the interventional channel 
\ctikzfig{thm-8-causal-effect}
from $X$ to $C$ is identifiable.
\end{prop}

\begingroup
\allowdisplaybreaks
\begin{proof}
First, we compute the generalized matrix elements of $g$, similarly to before:
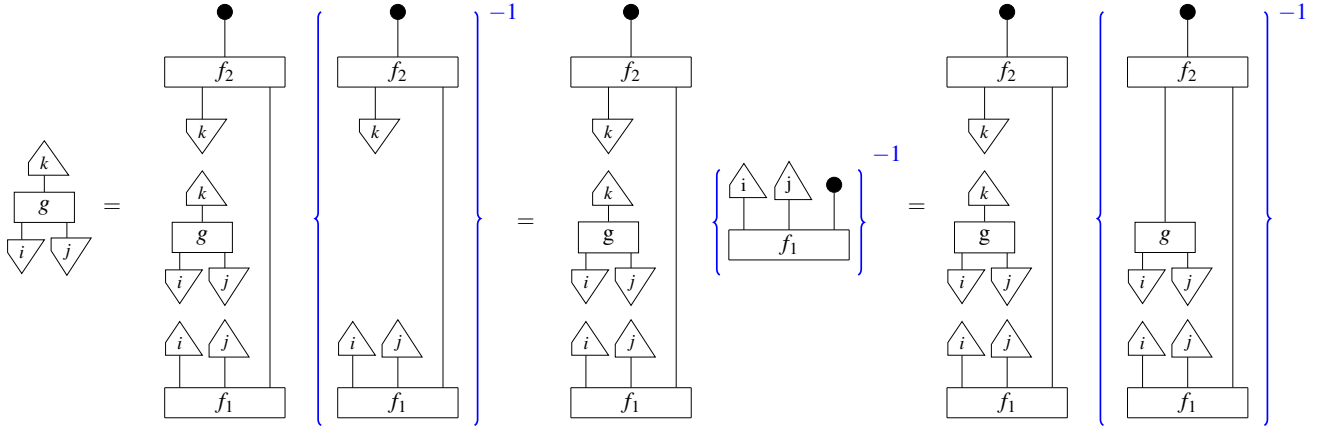
\[
    \input{g.tikz}
\]
Once we have the generalized matrix elements for $g$, we can again generate those of the outer comb by adjusting for $g$:
\ctikzfig{step2}
Finally, we compose the two processes at $A$ and $B$, leaving the $X$ input and output, to obtain the interventional channel.
\end{proof}
\endgroup

\section{Conclusion}
Functorial causal models, combined with string diagrammatic language, promise continued developments on multiple fronts of quantum and classical causal inference.  Since our comb factorization roughly amounts to Tian and Pearl's c-component factorization \cite{tian_general_2002}, we expect to be able to deal with more complicated quantum scenarios than the ones presented here, by porting classical identification protocols based on c-component factorization through our process-theoretic formalism to the category of quantum models.  Moreover, in both the quantum and classical settings, understanding causal inference as invoking a process theory's property of local process tomography unlocks the potential for immediately applying the abstract techniques of this article to inference with data from more general instruments than projective measurement or classical passive observation.  While on the classical side both sorts of generalization--to more complicated network shapes and to other data-collection instruments--may first lead simply to more efficient presentations of existing theory, all developments on the quantum side will constitute new domain knowledge.

Because graphs representing causal structure in other literature are often taken to encode stronger assumptions about complete common causes than are expressible in our framework, some identifiability conditions based on tests of such graphs do not have analogues in terms of the diagrams used in this article; our front-door criterion might be considered only a limited analogue of the classical set of sufficient conditions known by that name.  Our diagrams' lack of explicit representation of completeness of common causes is valuable in allowing us to discern which classical graphical criteria do not involve considerations of independence of multiple variables conditioned on observed complete common causes, and to derive the quantum analogues of those criteria without a treatment of quantum complete common causes.  Future work, however, will extend the framework here to incorporate assumptions of completeness of observed common causes, with a view to unified process-theoretic description of those parts of classical and quantum causal inference that rely on such assumptions.

\paragraph{Acknowledgements.} The authors would like to thank Rob Spekkens, Jon Barrett, Frederick Eberhardt, and Sally Shrapnel for useful discussions about causal structures, observation, common causes, and control of confounding. AK acknowledges support of Grant No. 61466 and No. 62312 from the John Templeton Foundation as part of the QISS project. The opinions expressed in this publication are those of the authors and do not necessarily reflect the views of the John Templeton Foundation.  IF acknowledges support from a Future of Humanity Institute DPhil Scholarship.

\bibliographystyle{eptcs}
\bibliography{id}

\end{document}

%% file: basic.tikzstyles
\tikzstyle{open node}=[fill=none, draw=black, shape=circle]
\tikzstyle{closed node}=[fill=black, draw=black, shape=circle, inner sep=0mm, minimum width=2.5mm]
\tikzstyle{medium box}=[fill=white, draw=black, shape=rectangle, minimum width=1cm, minimum height=0.4cm]
\tikzstyle{small box}=[fill=white, draw=black, shape=rectangle, minimum width=.6cm, minimum height=.4cm]
\tikzstyle{semilarge box}=[shape=rectangle, draw=black, fill=white, small box, minimum width=12.5mm, tikzit category=boxes]
\tikzstyle{left label}=[label, anchor=east, xshift=2mm, tikzit draw=green, tikzit fill=white, tikzit category=labels]
\tikzstyle{right label}=[label, anchor=west, xshift=-2mm, tikzit draw=purple, tikzit fill=white, tikzit category=labels]
\tikzstyle{state}=[fill=white, draw=black, shape=regular polygon, tikzit shape=rectangle, regular polygon sides=3, regular polygon rotate=60, scale=0.7, inner sep=1pt, tikzit fill=green]
\tikzstyle{NEbox}=[fill=white, draw=black, shape=NEbox]
\tikzstyle{NEtriangle}=[fill=white, draw=black, shape=NEtriangle, tikzit fill={rgb,255: red,255; green,116; blue,17}, scale=0.8, minimum height=5mm]
\tikzstyle{SEtriangle}=[fill=white, draw=black, shape=SEtriangle, tikzit fill={rgb,255: red,255; green,255; blue,183}, scale=0.8, minimum height=5mm]
\tikzstyle{effect}=[fill=white, draw=black, shape=regular polygon, tikzit shape=rectangle, regular polygon sides=3, regular polygon rotation=240, scale=0.7, inner sep=1pt, tikzit fill=red]

\tikzstyle{pointed}=[->]
\tikzstyle{thick wire}=[-, thick]
\tikzstyle{paren edge}=[-, tikzit draw=blue, thick, decorate, decoration={brace,amplitude=1mm,raise=-1mm}, draw=blue]

%% file: figures/causal.tikz
\begin{tikzpicture}
	\begin{pgfonlayer}{nodelayer}
		\node [style=small box] (0) at (-1.5, 0) {$f$};
		\node [style=closed node] (1) at (-1.5, 1.75) {};
		\node [style=none] (2) at (-1.5, -1.5) {};
		\node [style=right label] (3) at (-1.25, -1.25) {$A$};
		\node [style=right label] (4) at (-1.25, 1) {$B$};
		\node [style=none] (5) at (0.25, 0) {$=$};
		\node [style=none] (6) at (1.5, -0.75) {};
		\node [style=right label] (7) at (1.75, -0.25) {$A$};
		\node [style=closed node] (8) at (1.5, 0.75) {};
	\end{pgfonlayer}
	\begin{pgfonlayer}{edgelayer}
		\draw (2.center) to (0);
		\draw (0) to (1);
		\draw (6.center) to (8);
	\end{pgfonlayer}
\end{tikzpicture}

%% file: figures/line_yank.tikz
\begin{tikzpicture}
	\begin{pgfonlayer}{nodelayer}
		\node [style=none] (0) at (-5.5, -1.25) {};
		\node [style=none] (1) at (0, -1.25) {};
		\node [style=none] (2) at (-4, 0) {};
		\node [style=none] (3) at (-2.5, 0) {};
		\node [style=none] (4) at (-1.25, 0) {$=$};
		\node [style=none] (5) at (-5.5, 0) {};
		\node [style=none] (6) at (-4, 0) {};
		\node [style=none] (7) at (-2.5, 1.25) {};
		\node [style=none] (8) at (0, 1.25) {};
		\node [style=none] (13) at (5.5, -1.25) {};
		\node [style=none] (14) at (4, 0) {};
		\node [style=none] (15) at (2.5, 0) {};
		\node [style=none] (16) at (5.5, 0) {};
		\node [style=none] (17) at (4, 0) {};
		\node [style=none] (18) at (2.5, 1.25) {};
		\node [style=none] (22) at (1.25, 0) {$=$};
		\node [style=none] (24) at (4.25, 0.25) {};
	\end{pgfonlayer}
	\begin{pgfonlayer}{edgelayer}
		\draw [in=-90, out=-90, looseness=1.75] (2.center) to (3.center);
		\draw (3.center) to (7.center);
		\draw [in=90, out=90, looseness=1.75] (5.center) to (6.center);
		\draw (0.center) to (5.center);
		\draw [style=swap] (8.center) to (1.center);
		\draw [in=-90, out=-90, looseness=1.75] (14.center) to (15.center);
		\draw (15.center) to (18.center);
		\draw [in=90, out=90, looseness=1.75] (16.center) to (17.center);
		\draw (13.center) to (16.center);
	\end{pgfonlayer}
\end{tikzpicture}

%% file: figures/2-comb-ch.tikz
\begin{tikzpicture}
	\begin{pgfonlayer}{nodelayer}
		\node [style=semilarge box] (0) at (0, 0) {$f$};
		\node [style=none] (1) at (0.75, -0.5) {};
		\node [style=none] (2) at (0.75, -1.5) {};
		\node [style=right label] (3) at (1, -1) {$A'$};
		\node [style=none] (4) at (-0.75, 1.5) {};
		\node [style=none] (5) at (-0.75, 0.5) {};
		\node [style=right label] (6) at (-0.5, 1) {$A$};
		\node [style=none] (7) at (-0.75, -0.5) {};
		\node [style=none] (8) at (-0.75, -1.5) {};
		\node [style=right label] (9) at (-0.5, -1) {$B$};
		\node [style=none] (10) at (0.75, 1.5) {};
		\node [style=none] (11) at (0.75, 0.5) {};
		\node [style=right label] (12) at (1, 1) {$B'$};
	\end{pgfonlayer}
	\begin{pgfonlayer}{edgelayer}
		\draw (1.center) to (2.center);
		\draw (4.center) to (5.center);
		\draw (7.center) to (8.center);
		\draw (10.center) to (11.center);
	\end{pgfonlayer}
\end{tikzpicture}

%% file: figures/2-comb2.tikz
\begin{tikzpicture}
	\begin{pgfonlayer}{nodelayer}
		\node [style=none] (0) at (-1, 2.5) {};
		\node [style=none] (1) at (-1, 1.5) {};
		\node [style=none] (2) at (0.5, 1.5) {};
		\node [style=none] (3) at (0.5, -1.5) {};
		\node [style=none] (4) at (-1, -1.5) {};
		\node [style=none] (5) at (-1, -2.5) {};
		\node [style=none] (6) at (1.5, -2.5) {};
		\node [style=none] (7) at (1.5, 2.5) {};
		\node [style=none] (8) at (1, 0) {$f$};
		\node [style=none] (9) at (-0.25, -1) {};
		\node [style=none] (10) at (-0.25, -1.5) {};
		\node [style=none] (11) at (-0.25, -2.5) {};
		\node [style=none] (12) at (-0.25, -3) {};
		\node [style=none] (13) at (-0.25, 3) {};
		\node [style=none] (14) at (-0.25, 2.5) {};
		\node [style=none] (15) at (-0.25, 1.5) {};
		\node [style=none] (16) at (-0.25, 1) {};
		\node [style=left label] (17) at (-0.5, -3) {$B$};
		\node [style=left label] (18) at (-0.5, -1) {$A$};
		\node [style=left label] (19) at (-0.5, 1) {$A'$};
		\node [style=left label] (20) at (-0.5, 3) {$B'$};
	\end{pgfonlayer}
	\begin{pgfonlayer}{edgelayer}
		\draw (0.center) to (1.center);
		\draw (1.center) to (2.center);
		\draw (2.center) to (3.center);
		\draw (3.center) to (4.center);
		\draw (4.center) to (5.center);
		\draw (5.center) to (6.center);
		\draw (6.center) to (7.center);
		\draw (7.center) to (0.center);
		\draw (9.center) to (10.center);
		\draw (11.center) to (12.center);
		\draw (13.center) to (14.center);
		\draw (15.center) to (16.center);
	\end{pgfonlayer}
\end{tikzpicture}

%% file: figures/2-comb-compose.tikz
\begin{tikzpicture}
	\begin{pgfonlayer}{nodelayer}
		\node [style=none] (0) at (-4.25, 2.5) {};
		\node [style=none] (1) at (-4.25, 1.5) {};
		\node [style=none] (2) at (-2.5, 1.5) {};
		\node [style=none] (3) at (-2.5, -1.5) {};
		\node [style=none] (4) at (-4.25, -1.5) {};
		\node [style=none] (5) at (-4.25, -2.5) {};
		\node [style=none] (6) at (-1.5, -2.5) {};
		\node [style=none] (7) at (-1.5, 2.5) {};
		\node [style=none] (8) at (-2, 0) {$f$};
		\node [style=none] (9) at (-3.5, -0.5) {};
		\node [style=none] (10) at (-3.5, -1.5) {};
		\node [style=none] (11) at (-3.5, -2.5) {};
		\node [style=none] (12) at (-3.5, -3) {};
		\node [style=none] (13) at (-3.5, 3) {};
		\node [style=none] (14) at (-3.5, 2.5) {};
		\node [style=none] (15) at (-3.5, 1.5) {};
		\node [style=none] (16) at (-3.5, 0.5) {};
		\node [style=left label] (17) at (-3.75, -3) {$B$};
		\node [style=left label] (18) at (-3.75, -1) {$A$};
		\node [style=left label] (19) at (-3.75, 1) {$A'$};
		\node [style=left label] (20) at (-3.75, 3) {$B'$};
		\node [style=small box] (21) at (-3.5, 0) {$g$};
		\node [style=none] (22) at (0, 0) {$:=$};
		\node [style=semilarge box] (23) at (3, -1.5) {$f$};
		\node [style=none] (24) at (3.75, -2) {};
		\node [style=none] (25) at (3.75, -2.25) {};
		\node [style=left label] (26) at (3.5, -2.5) {$A'$};
		\node [style=none] (27) at (2.25, 0) {};
		\node [style=none] (28) at (2.25, -1) {};
		\node [style=none] (29) at (5.25, 1) {};
		\node [style=none] (30) at (2.25, 1) {};
		\node [style=left label] (31) at (2, -0.5) {$A$};
		\node [style=left label] (32) at (2, 1.5) {$A'$};
		\node [style=small box] (33) at (2.25, 0.5) {$g$};
		\node [style=none] (34) at (2.25, -2) {};
		\node [style=none] (35) at (2.25, -3) {};
		\node [style=left label] (36) at (2, -2.5) {$B$};
		\node [style=none] (37) at (3.75, 3) {};
		\node [style=none] (38) at (3.75, -1) {};
		\node [style=left label] (39) at (3.5, 2.5) {$B'$};
		\node [style=none] (40) at (5.25, -2.25) {};
	\end{pgfonlayer}
	\begin{pgfonlayer}{edgelayer}
		\draw (0.center) to (1.center);
		\draw (1.center) to (2.center);
		\draw (2.center) to (3.center);
		\draw (3.center) to (4.center);
		\draw (4.center) to (5.center);
		\draw (5.center) to (6.center);
		\draw (6.center) to (7.center);
		\draw (7.center) to (0.center);
		\draw (9.center) to (10.center);
		\draw (11.center) to (12.center);
		\draw (13.center) to (14.center);
		\draw (15.center) to (16.center);
		\draw (24.center) to (25.center);
		\draw (27.center) to (28.center);
		\draw [in=90, out=90] (29.center) to (30.center);
		\draw (34.center) to (35.center);
		\draw (37.center) to (38.center);
		\draw [in=-90, out=-90, looseness=1.50] (40.center) to (25.center);
		\draw (40.center) to (29.center);
	\end{pgfonlayer}
\end{tikzpicture}

%% file: figures/causal-model.tikz
\begin{tikzpicture}
	\begin{pgfonlayer}{nodelayer}
		\node [style=small box] (1) at (13, 0) {$c$};
		\node [style=semilarge box] (3) at (10.75, 0) {$b$};
		\node [style=NEtriangle] (4) at (9.75, -3) {$z$};
		\node [style=none] (5) at (10.75, 1.75) {};
		\node [style=none] (6) at (10.75, 3.75) {};
		\node [style=none] (7) at (10.75, 2.75) {};
		\node [style=none] (8) at (10.75, 0.25) {};
		\node [style=none] (9) at (9.75, -2) {};
		\node [style=none] (10) at (9.75, -1) {};
		\node [style=none] (11) at (9.75, -0.25) {};
		\node [style=semilarge box] (12) at (12.25, -3) {$a$};
		\node [style=none] (13) at (11.5, -2.75) {};
		\node [style=none] (14) at (11.5, -0.25) {};
		\node [style=none] (17) at (13, -2.75) {};
		\node [style=none] (18) at (13, -2) {};
		\node [style=none] (19) at (13, -1) {};
		\node [style=none] (20) at (13, 1.75) {};
		\node [style=none] (21) at (13, 3.75) {};
		\node [style=none] (22) at (13, 2.75) {};
		\node [style=right label] (23) at (10, -1.5) {$X_1$};
		\node [style=right label] (24) at (13.25, -1.5) {$X_2$};
		\node [style=right label] (25) at (11, 2.25) {$X_3$};
		\node [style=right label] (26) at (13.25, 2.25) {$X_4$};
		\node [style=semilarge box, minimum width=2cm] (27) at (11.875, 3.75) {$d$};
		\node [style=none] (52) at (-13.75, 0) {$:=$};
		\node [style=semilarge box, minimum width=2.8cm, minimum height=1.5cm] (53) at (-18.25, 0) {$D$};
		\node [style=none] (54) at (-20.5, -1.25) {};
		\node [style=none] (55) at (-20.5, -2.5) {};
		\node [style=none] (56) at (-20.5, 2.5) {};
		\node [style=none] (57) at (-20.5, 1.25) {};
		\node [style=right label] (58) at (-20.25, -2.25) {$X_1$};
		\node [style=right label] (59) at (-20.25, 2.25) {$X_1$};
		\node [style=none] (60) at (-19, -1.25) {};
		\node [style=none] (61) at (-19, -2.5) {};
		\node [style=none] (62) at (-19, 2.5) {};
		\node [style=none] (63) at (-19, 1.25) {};
		\node [style=right label] (64) at (-18.75, -2.25) {$X_2$};
		\node [style=right label] (65) at (-18.75, 2.25) {$X_2$};
		\node [style=none] (66) at (-17.5, -1.25) {};
		\node [style=none] (67) at (-17.5, -2.5) {};
		\node [style=none] (68) at (-17.5, 2.5) {};
		\node [style=none] (69) at (-17.5, 1.25) {};
		\node [style=right label] (70) at (-17.25, -2.25) {$X_3$};
		\node [style=right label] (71) at (-17.25, 2.25) {$X_3$};
		\node [style=none] (72) at (-16, -1.25) {};
		\node [style=none] (73) at (-16, -2.5) {};
		\node [style=none] (74) at (-16, 2.5) {};
		\node [style=none] (75) at (-16, 1.25) {};
		\node [style=right label] (76) at (-15.75, -2.25) {$X_4$};
		\node [style=right label] (77) at (-15.75, 2.25) {$X_4$};
		\node [style=small box] (78) at (-4.75, -0.25) {$c$};
		\node [style=semilarge box] (79) at (-8.5, -0.25) {$b$};
		\node [style=NEtriangle] (80) at (-10.75, 2.25) {$z$};
		\node [style=none] (81) at (-8.5, 1) {};
		\node [style=none] (82) at (-2.75, -0.25) {};
		\node [style=none] (83) at (-2.75, -4.25) {};
		\node [style=none] (84) at (-8.5, 0) {};
		\node [style=none] (85) at (-10.75, 4.25) {};
		\node [style=none] (86) at (-9.5, -4.25) {};
		\node [style=none] (87) at (-9.5, -0.5) {};
		\node [style=semilarge box] (88) at (-7, -2.25) {$a$};
		\node [style=none] (89) at (-7.75, -2) {};
		\node [style=none] (90) at (-7.75, -0.5) {};
		\node [style=none] (91) at (-6.25, -2) {};
		\node [style=none] (92) at (-6.25, 1) {};
		\node [style=none] (93) at (-4.75, -4.25) {};
		\node [style=none] (94) at (-4.75, 4.25) {};
		\node [style=none] (95) at (-0.5, -0.25) {};
		\node [style=none] (96) at (-0.5, -4.25) {};
		\node [style=right label] (97) at (-9.25, -4.25) {$X_1$};
		\node [style=right label] (98) at (-4.5, -4.25) {$X_2$};
		\node [style=right label] (99) at (-2.5, -4.25) {$X_3$};
		\node [style=right label] (100) at (-0.25, -4.25) {$X_4$};
		\node [style=semilarge box, minimum width=2cm] (101) at (-1.625, -0.25) {$d$};
		\node [style=right label] (102) at (-10.5, 4.25) {$X_1$};
		\node [style=right label] (103) at (-8.25, 4.25) {$X_2$};
		\node [style=right label] (104) at (-6, 4.25) {$X_3$};
		\node [style=right label] (105) at (-4.5, 4.25) {$X_4$};
		\node [style=none] (106) at (-8.5, 3.25) {};
		\node [style=none] (107) at (-8.5, 4.25) {};
		\node [style=none] (108) at (-6.25, 3.25) {};
		\node [style=none] (109) at (-6.25, 4.25) {};
		\node [style=none] (110) at (5.5, 0) {$\leadsto$};
	\end{pgfonlayer}
	\begin{pgfonlayer}{edgelayer}
		\draw (7.center) to (6.center);
		\draw (8.center) to (5.center);
		\draw (4) to (9.center);
		\draw (10.center) to (11.center);
		\draw (13.center) to (14.center);
		\draw (19.center) to (1);
		\draw (17.center) to (18.center);
		\draw (22.center) to (21.center);
		\draw (1) to (20.center);
		\draw (55.center) to (54.center);
		\draw (57.center) to (56.center);
		\draw (61.center) to (60.center);
		\draw (63.center) to (62.center);
		\draw (67.center) to (66.center);
		\draw (69.center) to (68.center);
		\draw (73.center) to (72.center);
		\draw (75.center) to (74.center);
		\draw (83.center) to (82.center);
		\draw (84.center) to (81.center);
		\draw (80) to (85.center);
		\draw (86.center) to (87.center);
		\draw (89.center) to (90.center);
		\draw (93.center) to (78);
		\draw [in=270, out=90] (91.center) to (92.center);
		\draw (96.center) to (95.center);
		\draw (78) to (94.center);
		\draw [in=270, out=90] (92.center) to (106.center);
		\draw (106.center) to (107.center);
		\draw [in=270, out=90] (81.center) to (108.center);
		\draw (108.center) to (109.center);
	\end{pgfonlayer}
\end{tikzpicture}

%% file: figures/loc-proc-tomog.tikz
\begin{tikzpicture}
	\begin{pgfonlayer}{nodelayer}
		\node [style=medium box] (0) at (0, 0) {$f$};
		\node [style=none] (1) at (-0.75, 0.5) {};
		\node [style=none] (2) at (0.75, 0.5) {};
		\node [style=none] (3) at (-0.75, -0.5) {};
		\node [style=none] (4) at (0.75, -0.5) {};
		\node [style=none] (9) at (-1.25, -1.25) {$A$};
		\node [style=none] (10) at (0.25, -1.25) {$B$};
		\node [style=none] (11) at (-1.25, 1.25) {$C$};
		\node [style=none] (12) at (0.25, 1.25) {$D$};
		\node [style=SEtriangle] (13) at (-0.75, 2.5) {$k$};
		\node [style=SEtriangle] (14) at (0.75, 2.5) {$l$};
		\node [style=NEtriangle] (15) at (-0.75, -2.5) {$i$};
		\node [style=NEtriangle] (16) at (0.75, -2.5) {$j$};
	\end{pgfonlayer}
	\begin{pgfonlayer}{edgelayer}
		\draw (3.center) to (15);
		\draw (4.center) to (16);
		\draw (2.center) to (14);
		\draw (1.center) to (13);
	\end{pgfonlayer}
\end{tikzpicture}

%% file: figures/int-outcome.tikz
\begin{tikzpicture}
	\begin{pgfonlayer}{nodelayer}
		\node [style=SEtriangle] (0) at (0, -1.25) {$i$};
		\node [style=NEtriangle] (1) at (0, 1.25) {$j$};
		\node [style=none] (2) at (0, -2.75) {};
		\node [style=none] (3) at (0, 2.75) {};
	\end{pgfonlayer}
	\begin{pgfonlayer}{edgelayer}
		\draw (3.center) to (1);
		\draw (2.center) to (0);
	\end{pgfonlayer}
\end{tikzpicture}

%% file: figures/front-door-model.tikz
\begin{tikzpicture}
	\begin{pgfonlayer}{nodelayer}
		\node [style=none] (0) at (-2, -2.25) {};
		\node [style=none] (1) at (-2, -3) {};
		\node [style=none] (2) at (2, -3) {};
		\node [style=none] (3) at (2, -2.25) {};
		\node [style=none] (4) at (-1, -2.25) {};
		\node [style=none] (6) at (-1, -1.5) {};
		\node [style=none] (7) at (-1, -0.75) {};
		\node [style=none] (9) at (-1, 0.75) {};
		\node [style=none] (10) at (-1, 1.5) {};
		\node [style=none] (11) at (-1, 2.25) {};
		\node [style=none] (12) at (-2, 2.25) {};
		\node [style=none] (13) at (-2, 3) {};
		\node [style=none] (14) at (2, 3) {};
		\node [style=none] (15) at (2, 2.25) {};
		\node [style=none] (16) at (1, 2.25) {};
		\node [style=none] (17) at (1, -2.25) {};
		\node [style=none] (20) at (0, 2.5) {y};
		\node [style=none] (21) at (0, -2.75) {u};
		\node [style=none] (22) at (0, 3.75) {};
		\node [style=none] (23) at (0, 4.5) {};
		\node [style=closed node] (24) at (0, 5.25) {};
		\node [style=none] (25) at (0, 3) {};
		\node [style=none] (26) at (-0.5, 4) {$Y$};
		\node [style=none] (27) at (-1.5, -1.25) {$X$};
		\node [style=none] (28) at (-1.5, 1) {$Z$};
		\node [style=small box] (30) at (-1, 0) {z};
	\end{pgfonlayer}
	\begin{pgfonlayer}{edgelayer}
		\draw (13.center) to (14.center);
		\draw (14.center) to (15.center);
		\draw (13.center) to (12.center);
		\draw (12.center) to (15.center);
		\draw (11.center) to (10.center);
		\draw (16.center) to (17.center);
		\draw (17.center) to (3.center);
		\draw (3.center) to (2.center);
		\draw (2.center) to (1.center);
		\draw (1.center) to (0.center);
		\draw (0.center) to (4.center);
		\draw (4.center) to (17.center);
		\draw (25.center) to (22.center);
		\draw (23.center) to (24);
		\draw (7.center) to (30);
		\draw (30) to (9.center);
		\draw (6.center) to (4.center);
	\end{pgfonlayer}
\end{tikzpicture}

%% file: figures/z.tikz
\begin{tikzpicture}
	\begin{pgfonlayer}{nodelayer}
		\node [style=none] (0) at (1, -5.25) {};
		\node [style=none] (1) at (1, -6) {};
		\node [style=none] (2) at (5, -6) {};
		\node [style=none] (3) at (5, -5.25) {};
		\node [style=none] (4) at (2, -5.25) {};
		\node [style=none] (5) at (2, 3.25) {};
		\node [style=none] (6) at (1, 3.25) {};
		\node [style=none] (7) at (1, 4) {};
		\node [style=none] (8) at (5, 4) {};
		\node [style=none] (9) at (5, 3.25) {};
		\node [style=none] (10) at (4, 3.25) {};
		\node [style=none] (11) at (4, -5.25) {};
		\node [style=none] (12) at (3, 3.575) {$y$};
		\node [style=none] (13) at (3, -5.625) {$u$};
		\node [style=closed node] (14) at (3, 5.5) {};
		\node [style=none] (15) at (3, 4) {};
		\node [style=small box] (20) at (2, -1) {$z$};
		\node [style=NEtriangle] (21) at (2, 2.25) {$j$};
		\node [style=NEtriangle] (22) at (2, -2.25) {$i$};
		\node [style=SEtriangle] (23) at (2, 0.25) {$j$};
		\node [style=SEtriangle] (24) at (2, -4.25) {$i$};
		\node [style=small box] (25) at (-2, 0) {$z$};
		\node [style=NEtriangle] (26) at (-2, -1.25) {$i$};
		\node [style=SEtriangle] (27) at (-2, 1.25) {$j$};
		\node [style=none] (28) at (7, -5.25) {};
		\node [style=none] (29) at (7, -6) {};
		\node [style=none] (30) at (11, -6) {};
		\node [style=none] (31) at (11, -5.25) {};
		\node [style=none] (32) at (8, -5.25) {};
		\node [style=none] (33) at (8, 3.25) {};
		\node [style=none] (34) at (7, 3.25) {};
		\node [style=none] (35) at (7, 4) {};
		\node [style=none] (36) at (11, 4) {};
		\node [style=none] (37) at (11, 3.25) {};
		\node [style=none] (38) at (10, 3.25) {};
		\node [style=none] (39) at (10, -5.25) {};
		\node [style=none] (40) at (9, 3.575) {$y$};
		\node [style=none] (41) at (9, -5.625) {$u$};
		\node [style=closed node] (42) at (9, 5.5) {};
		\node [style=none] (43) at (9, 4) {};
		\node [style=NEtriangle] (46) at (8, 2.25) {$j$};
		\node [style=SEtriangle] (47) at (8, -4.25) {$i$};
		\node [style=none] (48) at (12, 5.5) {\color{blue}$-1$};
		\node [style=none] (49) at (0, 0) {$=$};
		\node [style=none] (50) at (14.25, -5.25) {};
		\node [style=none] (51) at (14.25, -6) {};
		\node [style=none] (52) at (18.25, -6) {};
		\node [style=none] (53) at (18.25, -5.25) {};
		\node [style=none] (54) at (15.25, -5.25) {};
		\node [style=none] (55) at (15.25, 3.25) {};
		\node [style=none] (56) at (14.25, 3.25) {};
		\node [style=none] (57) at (14.25, 4) {};
		\node [style=none] (58) at (18.25, 4) {};
		\node [style=none] (59) at (18.25, 3.25) {};
		\node [style=none] (60) at (17.25, 3.25) {};
		\node [style=none] (61) at (17.25, -5.25) {};
		\node [style=none] (62) at (16.25, 3.575) {$y$};
		\node [style=none] (63) at (16.25, -5.625) {$u$};
		\node [style=closed node] (64) at (16.25, 5.5) {};
		\node [style=none] (65) at (16.25, 4) {};
		\node [style=small box] (70) at (15.25, -1) {$z$};
		\node [style=NEtriangle] (71) at (15.25, 2.25) {$j$};
		\node [style=NEtriangle] (72) at (15.25, -2.25) {$i$};
		\node [style=SEtriangle] (73) at (15.25, 0.25) {$j$};
		\node [style=SEtriangle] (74) at (15.25, -4.25) {$i$};
		\node [style=none] (75) at (19, -1) {};
		\node [style=none] (76) at (19, -1.75) {};
		\node [style=none] (77) at (23, -1.75) {};
		\node [style=none] (78) at (23, -1) {};
		\node [style=none] (79) at (20, -1) {};
		\node [style=none] (80) at (22, -1) {};
		\node [style=none] (81) at (21, -1.375) {$u$};
		\node [style=closed node] (82) at (22, 1) {};
		\node [style=SEtriangle] (84) at (20, 0) {$i$};
		\node [style=none] (86) at (13.25, -1) {$=$};
		\node [style=none] (87) at (6.5, -6.5) {};
		\node [style=none] (88) at (6.5, 5.5) {};
		\node [style=none] (96) at (26.25, -5.5) {};
		\node [style=none] (97) at (26.25, -6.25) {};
		\node [style=none] (98) at (30.25, -6.25) {};
		\node [style=none] (99) at (30.25, -5.5) {};
		\node [style=none] (100) at (27.25, -5.5) {};
		\node [style=none] (101) at (27.25, 3) {};
		\node [style=none] (102) at (26.25, 3) {};
		\node [style=none] (103) at (26.25, 3.75) {};
		\node [style=none] (104) at (30.25, 3.75) {};
		\node [style=none] (105) at (30.25, 3) {};
		\node [style=none] (106) at (29.25, 3) {};
		\node [style=none] (107) at (29.25, -5.5) {};
		\node [style=none] (108) at (28.25, 3.3) {$y$};
		\node [style=none] (109) at (28.25, -5.875) {$u$};
		\node [style=closed node] (110) at (28.25, 5.25) {};
		\node [style=none] (111) at (28.25, 3.75) {};
		\node [style=small box] (116) at (27.25, -1.25) {$z$};
		\node [style=NEtriangle] (117) at (27.25, 2) {$j$};
		\node [style=NEtriangle] (118) at (27.25, -2.5) {$i$};
		\node [style=SEtriangle] (119) at (27.25, 0) {$j$};
		\node [style=SEtriangle] (120) at (27.25, -4.5) {$i$};
		\node [style=none] (121) at (25.25, 0) {$=$};
		\node [style=none] (122) at (31.75, -4.75) {};
		\node [style=none] (123) at (31.75, -5.5) {};
		\node [style=none] (124) at (35.75, -5.5) {};
		\node [style=none] (125) at (35.75, -4.75) {};
		\node [style=none] (126) at (32.75, -4.75) {};
		\node [style=none] (127) at (32.75, 2.25) {};
		\node [style=none] (128) at (31.75, 2.25) {};
		\node [style=none] (129) at (31.75, 3) {};
		\node [style=none] (130) at (35.75, 3) {};
		\node [style=none] (131) at (35.75, 2.25) {};
		\node [style=none] (132) at (34.75, 2.25) {};
		\node [style=none] (133) at (34.75, -4.75) {};
		\node [style=none] (134) at (33.75, 2.55) {$y$};
		\node [style=none] (135) at (33.75, -5.125) {$u$};
		\node [style=closed node] (136) at (33.75, 4.5) {};
		\node [style=none] (137) at (33.75, 3) {};
		\node [style=small box] (142) at (32.75, -0.25) {$z$};
		\node [style=NEtriangle] (143) at (32.75, -1.75) {$i$};
		\node [style=SEtriangle] (144) at (32.75, -3.75) {$i$};
		\node [style=none] (151) at (11.5, 5.5) {};
		\node [style=none] (152) at (11.5, -6.5) {};
		\node [style=none] (153) at (24, 1.25) {\color{blue}$-1$};
		\node [style=none] (154) at (18.5, -2.25) {};
		\node [style=none] (155) at (18.5, 1.25) {};
		\node [style=none] (156) at (23.5, 1.25) {};
		\node [style=none] (157) at (23.5, -2.25) {};
		\node [style=none] (158) at (36.75, 4.75) {\color{blue}$-1$};
		\node [style=none] (159) at (31.25, -6) {};
		\node [style=none] (160) at (31.25, 4.75) {};
		\node [style=none] (161) at (36.25, 4.75) {};
		\node [style=none] (162) at (36.25, -6) {};
	\end{pgfonlayer}
	\begin{pgfonlayer}{edgelayer}
		\draw (7.center) to (8.center);
		\draw (8.center) to (9.center);
		\draw (7.center) to (6.center);
		\draw (6.center) to (9.center);
		\draw (10.center) to (11.center);
		\draw (11.center) to (3.center);
		\draw (3.center) to (2.center);
		\draw (2.center) to (1.center);
		\draw (1.center) to (0.center);
		\draw (0.center) to (4.center);
		\draw (4.center) to (11.center);
		\draw (22) to (20);
		\draw (23) to (20);
		\draw (5.center) to (21);
		\draw (14) to (15.center);
		\draw (26) to (25);
		\draw (27) to (25);
		\draw (35.center) to (36.center);
		\draw (36.center) to (37.center);
		\draw (35.center) to (34.center);
		\draw (34.center) to (37.center);
		\draw (38.center) to (39.center);
		\draw (39.center) to (31.center);
		\draw (31.center) to (30.center);
		\draw (30.center) to (29.center);
		\draw (29.center) to (28.center);
		\draw (28.center) to (32.center);
		\draw (32.center) to (39.center);
		\draw (33.center) to (46);
		\draw (42) to (43.center);
		\draw (57.center) to (58.center);
		\draw (58.center) to (59.center);
		\draw (57.center) to (56.center);
		\draw (56.center) to (59.center);
		\draw (60.center) to (61.center);
		\draw (61.center) to (53.center);
		\draw (53.center) to (52.center);
		\draw (52.center) to (51.center);
		\draw (51.center) to (50.center);
		\draw (50.center) to (54.center);
		\draw (54.center) to (61.center);
		\draw (72) to (70);
		\draw (73) to (70);
		\draw (55.center) to (71);
		\draw (64) to (65.center);
		\draw (80.center) to (78.center);
		\draw (78.center) to (77.center);
		\draw (77.center) to (76.center);
		\draw (76.center) to (75.center);
		\draw (75.center) to (79.center);
		\draw (79.center) to (80.center);
		\draw (80.center) to (82);
		\draw (103.center) to (104.center);
		\draw (104.center) to (105.center);
		\draw (103.center) to (102.center);
		\draw (102.center) to (105.center);
		\draw (106.center) to (107.center);
		\draw (107.center) to (99.center);
		\draw (99.center) to (98.center);
		\draw (98.center) to (97.center);
		\draw (97.center) to (96.center);
		\draw (96.center) to (100.center);
		\draw (100.center) to (107.center);
		\draw (118) to (116);
		\draw (119) to (116);
		\draw (101.center) to (117);
		\draw (110) to (111.center);
		\draw (129.center) to (130.center);
		\draw (130.center) to (131.center);
		\draw (129.center) to (128.center);
		\draw (128.center) to (131.center);
		\draw (132.center) to (133.center);
		\draw (133.center) to (125.center);
		\draw (125.center) to (124.center);
		\draw (124.center) to (123.center);
		\draw (123.center) to (122.center);
		\draw (122.center) to (126.center);
		\draw (126.center) to (133.center);
		\draw (143) to (142);
		\draw (136) to (137.center);
		\draw (127.center) to (142);
		\draw [style=paren edge] (87.center) to (88.center);
		\draw [style=paren edge] (151.center) to (152.center);
		\draw [style=paren edge] (154.center) to (155.center);
		\draw [style=paren edge] (156.center) to (157.center);
		\draw [style=paren edge] (159.center) to (160.center);
		\draw [style=paren edge] (161.center) to (162.center);
		\draw (24) to (4.center);
		\draw (47) to (32.center);
		\draw (74) to (54.center);
		\draw (84) to (79.center);
		\draw (120) to (100.center);
		\draw (144) to (126.center);
	\end{pgfonlayer}
\end{tikzpicture}

%% file: figures/g.tikz
\begin{tikzpicture}
	\begin{pgfonlayer}{nodelayer}
		\node [style=medium box] (0) at (-14.25, 0) {$g$};
		\node [style=NEtriangle] (1) at (-15, -1.5) {$i$};
		\node [style=none] (2) at (-15, -0.25) {};
		\node [style=NEtriangle] (3) at (-13.5, -1.5) {$j$};
		\node [style=none] (4) at (-13.5, -0.25) {};
		\node [style=SEtriangle] (5) at (-14.25, 1.5) {$k$};
		\node [style=none] (6) at (-14.25, 0) {};
		\node [style=none] (7) at (-12, 0) {$=$};
		\node [style=none] (8) at (-10.25, -6) {};
		\node [style=none] (9) at (-10.25, -7) {};
		\node [style=none] (10) at (-6.25, -7) {};
		\node [style=none] (11) at (-6.25, -6) {};
		\node [style=none] (12) at (-8.25, -6) {};
		\node [style=none] (13) at (-8.25, -4.5) {};
		\node [style=none] (14) at (-9, 2.5) {};
		\node [style=none] (15) at (-9, 4) {};
		\node [style=none] (16) at (-10.25, 4) {};
		\node [style=none] (17) at (-10.25, 5) {};
		\node [style=none] (18) at (-6.25, 5) {};
		\node [style=none] (19) at (-6.25, 4) {};
		\node [style=none] (20) at (-6.75, 4) {};
		\node [style=none] (21) at (-6.75, -6) {};
		\node [style=none] (22) at (-8.25, 4.5) {$f_2$};
		\node [style=none] (23) at (-8.25, -6.5) {$f_1$};
		\node [style=none] (25) at (-8.25, 5) {};
		\node [style=closed node] (26) at (-8.25, 6.5) {};
		\node [style=none] (31) at (-9.75, -6) {};
		\node [style=none] (34) at (-9.75, -1.25) {};
		\node [style=none] (35) at (-8.25, -2.5) {};
		\node [style=none] (36) at (-8.25, -1.25) {};
		\node [style=medium box] (37) at (-9, -1) {$g$};
		\node [style=NEtriangle] (41) at (-9.75, -2.5) {$i$};
		\node [style=NEtriangle] (42) at (-8.25, -2.5) {$j$};
		\node [style=SEtriangle] (43) at (-9, 0.5) {$k$};
		\node [style=NEtriangle] (44) at (-9, 2.5) {$k$};
		\node [style=SEtriangle] (45) at (-9.75, -4.5) {$i$};
		\node [style=SEtriangle] (46) at (-8.25, -4.5) {$j$};
		\node [style=none] (74) at (1, 6.5) {\color{blue}$-1$};
		\node [style=none] (75) at (1.75, -0.5) {$=$};
		\node [style=none] (76) at (8.5, -0.75) {};
		\node [style=none] (77) at (8.5, -1.75) {};
		\node [style=none] (78) at (12.5, -1.75) {};
		\node [style=none] (79) at (12.5, -0.75) {};
		\node [style=none] (80) at (10.5, -0.75) {};
		\node [style=none] (82) at (12, -0.75) {};
		\node [style=none] (83) at (10.5, -1.25) {$f_1$};
		\node [style=none] (85) at (9, -0.75) {};
		\node [style=SEtriangle] (87) at (9, 0.75) {i};
		\node [style=SEtriangle] (88) at (10.5, 0.75) {j};
		\node [style=closed node] (89) at (12, 0.75) {};
		\node [style=none] (90) at (3.25, -6) {};
		\node [style=none] (91) at (3.25, -7) {};
		\node [style=none] (92) at (7.25, -7) {};
		\node [style=none] (93) at (7.25, -6) {};
		\node [style=none] (94) at (5.25, -6) {};
		\node [style=none] (97) at (4.5, 4) {};
		\node [style=none] (98) at (3.25, 4) {};
		\node [style=none] (99) at (3.25, 5) {};
		\node [style=none] (100) at (7.25, 5) {};
		\node [style=none] (101) at (7.25, 4) {};
		\node [style=none] (102) at (6.75, 4) {};
		\node [style=none] (103) at (6.75, -6) {};
		\node [style=none] (104) at (5.25, 4.5) {$f_2$};
		\node [style=none] (105) at (5.25, -6.5) {$f_1$};
		\node [style=none] (107) at (5.25, 5) {};
		\node [style=closed node] (108) at (5.25, 6.5) {};
		\node [style=none] (113) at (3.75, -6) {};
		\node [style=none] (116) at (3.75, -1.25) {};
		\node [style=none] (118) at (5.25, -1.25) {};
		\node [style=medium box] (119) at (4.5, -1) {g};
		\node [style=NEtriangle] (123) at (3.75, -2.5) {$i$};
		\node [style=NEtriangle] (124) at (5.25, -2.5) {$j$};
		\node [style=SEtriangle] (125) at (4.5, 0.5) {$k$};
		\node [style=NEtriangle] (126) at (4.5, 2.5) {$k$};
		\node [style=SEtriangle] (127) at (3.75, -4.5) {$i$};
		\node [style=SEtriangle] (128) at (5.25, -4.5) {$j$};
		\node [style=none] (130) at (-4.5, -6) {};
		\node [style=none] (131) at (-4.5, -7) {};
		\node [style=none] (132) at (-0.5, -7) {};
		\node [style=none] (133) at (-0.5, -6) {};
		\node [style=none] (134) at (-2.5, -6) {};
		\node [style=none] (135) at (-2.5, -4.5) {};
		\node [style=none] (136) at (-3.25, 2.5) {};
		\node [style=none] (137) at (-3.25, 4) {};
		\node [style=none] (138) at (-4.5, 4) {};
		\node [style=none] (139) at (-4.5, 5) {};
		\node [style=none] (140) at (-0.5, 5) {};
		\node [style=none] (141) at (-0.5, 4) {};
		\node [style=none] (142) at (-1, 4) {};
		\node [style=none] (143) at (-1, -6) {};
		\node [style=none] (144) at (-2.5, 4.5) {$f_2$};
		\node [style=none] (145) at (-2.5, -6.5) {$f_1$};
		\node [style=none] (146) at (-2.5, 5) {};
		\node [style=closed node] (147) at (-2.5, 6.5) {};
		\node [style=none] (151) at (-4, -6) {};
		\node [style=NEtriangle] (161) at (-3.25, 2.5) {$k$};
		\node [style=SEtriangle] (162) at (-4, -4.5) {$i$};
		\node [style=SEtriangle] (163) at (-2.5, -4.5) {$j$};
		\node [style=none] (164) at (-5.25, -7.25) {};
		\node [style=none] (165) at (-5.25, 6.5) {};
		\node [style=none] (166) at (0.25, 6.5) {};
		\node [style=none] (167) at (0.25, -7.25) {};
		\node [style=none] (168) at (13.75, 1.75) {\color{blue}$-1$};
		\node [style=none] (169) at (13, 1.25) {};
		\node [style=none] (170) at (13, -2.25) {};
		\node [style=none] (171) at (8, -2.25) {};
		\node [style=none] (172) at (8, 1.25) {};
		\node [style=none] (174) at (14.75, 0) {$=$};
		\node [style=none] (214) at (21.75, -6) {};
		\node [style=none] (215) at (21.75, -7) {};
		\node [style=none] (216) at (25.75, -7) {};
		\node [style=none] (217) at (25.75, -6) {};
		\node [style=none] (218) at (23.75, -6) {};
		\node [style=none] (220) at (23, 4) {};
		\node [style=none] (221) at (21.75, 4) {};
		\node [style=none] (222) at (21.75, 5) {};
		\node [style=none] (223) at (25.75, 5) {};
		\node [style=none] (224) at (25.75, 4) {};
		\node [style=none] (225) at (25.25, 4) {};
		\node [style=none] (226) at (25.25, -6) {};
		\node [style=none] (227) at (23.75, 4.5) {$f_2$};
		\node [style=none] (228) at (23.75, -6.5) {$f_1$};
		\node [style=none] (230) at (23.75, 5) {};
		\node [style=closed node] (231) at (23.75, 6.5) {};
		\node [style=none] (236) at (22.25, -6) {};
		\node [style=none] (239) at (22.25, -1.25) {};
		\node [style=none] (241) at (23.75, -1.25) {};
		\node [style=medium box] (242) at (23, -1) {$g$};
		\node [style=NEtriangle] (245) at (22.25, -2.5) {$i$};
		\node [style=NEtriangle] (246) at (23.75, -2.5) {$j$};
		\node [style=SEtriangle] (247) at (22.25, -4.5) {$i$};
		\node [style=SEtriangle] (248) at (23.75, -4.5) {$j$};
		\node [style=none] (249) at (27.25, 6.5) {\color{blue}$-1$};
		\node [style=none] (250) at (20.75, -7.25) {};
		\node [style=none] (251) at (20.75, 6.5) {};
		\node [style=none] (252) at (26.5, 6.5) {};
		\node [style=none] (253) at (26.5, -7.25) {};
		\node [style=none] (254) at (15.75, -6) {};
		\node [style=none] (255) at (15.75, -7) {};
		\node [style=none] (256) at (19.75, -7) {};
		\node [style=none] (257) at (19.75, -6) {};
		\node [style=none] (258) at (17.75, -6) {};
		\node [style=none] (259) at (17, 4) {};
		\node [style=none] (260) at (15.75, 4) {};
		\node [style=none] (261) at (15.75, 5) {};
		\node [style=none] (262) at (19.75, 5) {};
		\node [style=none] (263) at (19.75, 4) {};
		\node [style=none] (264) at (19.25, 4) {};
		\node [style=none] (265) at (19.25, -6) {};
		\node [style=none] (266) at (17.75, 4.5) {$f_2$};
		\node [style=none] (267) at (17.75, -6.5) {$f_1$};
		\node [style=none] (268) at (17.75, 5) {};
		\node [style=closed node] (269) at (17.75, 6.5) {};
		\node [style=none] (273) at (16.25, -6) {};
		\node [style=none] (274) at (16.25, -1.25) {};
		\node [style=none] (275) at (17.75, -1.25) {};
		\node [style=medium box] (276) at (17, -1) {g};
		\node [style=NEtriangle] (279) at (16.25, -2.5) {$i$};
		\node [style=NEtriangle] (280) at (17.75, -2.5) {$j$};
		\node [style=SEtriangle] (281) at (17, 0.5) {$k$};
		\node [style=NEtriangle] (282) at (17, 2.5) {$k$};
		\node [style=SEtriangle] (283) at (16.25, -4.5) {$i$};
		\node [style=SEtriangle] (284) at (17.75, -4.5) {$j$};
	\end{pgfonlayer}
	\begin{pgfonlayer}{edgelayer}
		\draw (2.center) to (1);
		\draw (4.center) to (3);
		\draw (6.center) to (5);
		\draw (17.center) to (18.center);
		\draw (18.center) to (19.center);
		\draw (17.center) to (16.center);
		\draw (16.center) to (19.center);
		\draw (15.center) to (14.center);
		\draw (20.center) to (21.center);
		\draw (21.center) to (11.center);
		\draw (11.center) to (10.center);
		\draw (10.center) to (9.center);
		\draw (9.center) to (8.center);
		\draw (8.center) to (12.center);
		\draw (12.center) to (21.center);
		\draw (25.center) to (26);
		\draw (35.center) to (36.center);
		\draw (82.center) to (79.center);
		\draw (79.center) to (78.center);
		\draw (78.center) to (77.center);
		\draw (77.center) to (76.center);
		\draw (76.center) to (80.center);
		\draw (80.center) to (82.center);
		\draw (89) to (82.center);
		\draw (99.center) to (100.center);
		\draw (100.center) to (101.center);
		\draw (99.center) to (98.center);
		\draw (98.center) to (101.center);
		\draw (102.center) to (103.center);
		\draw (103.center) to (93.center);
		\draw (93.center) to (92.center);
		\draw (92.center) to (91.center);
		\draw (91.center) to (90.center);
		\draw (90.center) to (94.center);
		\draw (94.center) to (103.center);
		\draw (107.center) to (108);
		\draw (41) to (34.center);
		\draw (31.center) to (45);
		\draw (37) to (43);
		\draw (123) to (116.center);
		\draw (113.center) to (127);
		\draw (124) to (118.center);
		\draw (119) to (125);
		\draw (97.center) to (126);
		\draw (139.center) to (140.center);
		\draw (140.center) to (141.center);
		\draw (139.center) to (138.center);
		\draw (138.center) to (141.center);
		\draw (137.center) to (136.center);
		\draw (142.center) to (143.center);
		\draw (143.center) to (133.center);
		\draw (133.center) to (132.center);
		\draw (132.center) to (131.center);
		\draw (131.center) to (130.center);
		\draw (130.center) to (134.center);
		\draw (134.center) to (143.center);
		\draw (146.center) to (147);
		\draw (151.center) to (162);
		\draw [style=paren edge] (164.center) to (165.center);
		\draw [style=paren edge] (166.center) to (167.center);
		\draw [style=paren edge] (169.center) to (170.center);
		\draw [style=paren edge] (171.center) to (172.center);
		\draw (85.center) to (87);
		\draw (222.center) to (223.center);
		\draw (223.center) to (224.center);
		\draw (222.center) to (221.center);
		\draw (221.center) to (224.center);
		\draw (225.center) to (226.center);
		\draw (226.center) to (217.center);
		\draw (217.center) to (216.center);
		\draw (216.center) to (215.center);
		\draw (215.center) to (214.center);
		\draw (214.center) to (218.center);
		\draw (218.center) to (226.center);
		\draw (230.center) to (231);
		\draw (220.center) to (242);
		\draw [style=paren edge] (250.center) to (251.center);
		\draw [style=paren edge] (252.center) to (253.center);
		\draw (261.center) to (262.center);
		\draw (262.center) to (263.center);
		\draw (261.center) to (260.center);
		\draw (260.center) to (263.center);
		\draw (264.center) to (265.center);
		\draw (265.center) to (257.center);
		\draw (257.center) to (256.center);
		\draw (256.center) to (255.center);
		\draw (255.center) to (254.center);
		\draw (254.center) to (258.center);
		\draw (258.center) to (265.center);
		\draw (268.center) to (269);
		\draw (279) to (274.center);
		\draw (273.center) to (283);
		\draw (280) to (275.center);
		\draw (276) to (281);
		\draw (259.center) to (282);
		\draw (245) to (239.center);
		\draw (246) to (241.center);
		\draw (236.center) to (247);
		\draw (163) to (134.center);
		\draw (128) to (94.center);
		\draw (284) to (258.center);
		\draw (248) to (218.center);
		\draw (88) to (80.center);
		\draw (46) to (12.center);
	\end{pgfonlayer}
\end{tikzpicture}